\documentclass[8.5pt,twoside,twocolumn]{article}
\usepackage[super,sort&compress,comma]{natbib} 
\usepackage{mhchem}
\usepackage{times,mathptmx}
\usepackage{sectsty}
\usepackage{balance} 

\usepackage{graphicx} 
\usepackage{lastpage}
\usepackage[format=plain,justification=raggedright,singlelinecheck=false,font=small,labelfont=bf,labelsep=space]{caption} 
\usepackage{fancyhdr}
\begin{document}

\thispagestyle{plain}
\fancypagestyle{plain}{
\fancyhead[L]{\includegraphics[height=8pt]{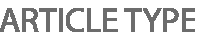}}
\fancyhead[C]{\hspace{-1cm}\includegraphics[height=20pt]{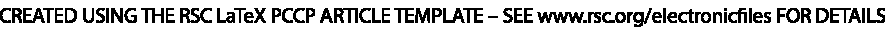}}
\fancyhead[R]{\includegraphics[height=10pt]{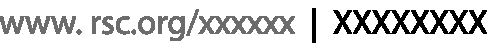}\vspace{-0.2cm}}
\renewcommand{\headrulewidth}{1pt}}
\renewcommand{\thefootnote}{\fnsymbol{footnote}}
\renewcommand\footnoterule{\vspace*{1pt}%
\hrule width 3.4in height 0.4pt \vspace*{5pt}} 
\setcounter{secnumdepth}{5}

\makeatletter 
\def\subsubsection{\@startsection{subsubsection}{3}{10pt}{-1.25ex plus -1ex minus -.1ex}{0ex plus 0ex}{\normalsize\bf}} 
\def\paragraph{\@startsection{paragraph}{4}{10pt}{-1.25ex plus -1ex minus -.1ex}{0ex plus 0ex}{\normalsize\textit}} 
\renewcommand\@biblabel[1]{#1}            
\renewcommand\@makefntext[1]%
{\noindent\makebox[0pt][r]{\@thefnmark\,}#1}
\makeatother 
\renewcommand{\figurename}{\small{Fig.}~}
\sectionfont{\large}
\subsectionfont{\normalsize} 

\fancyfoot{}
\fancyfoot[LO,RE]{\vspace{-7pt}\includegraphics[height=9pt]{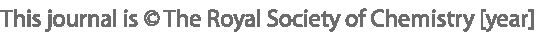}}
\fancyfoot[CO]{\vspace{-7.2pt}\hspace{12.2cm}\includegraphics{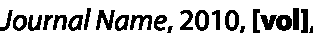}}
\fancyfoot[CE]{\vspace{-7.5pt}\hspace{-13.5cm}\includegraphics{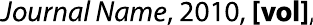}}
\fancyfoot[RO]{\footnotesize{\sffamily{1--\pageref{LastPage} ~\textbar  \hspace{2pt}\thepage}}}
\fancyfoot[LE]{\footnotesize{\sffamily{\thepage~\textbar\hspace{3.45cm} 1--\pageref{LastPage}}}}
\fancyhead{}
\renewcommand{\headrulewidth}{1pt} 
\renewcommand{\footrulewidth}{1pt}
\setlength{\arrayrulewidth}{1pt}
\setlength{\columnsep}{6.5mm}
\setlength\bibsep{1pt}

\twocolumn[
  \begin{@twocolumnfalse}
\noindent\LARGE{\textbf{Ab initio Investigation of Adsorption Characteristics of Bisphosphonates on Hydroxyapatite (001) Surface}}
\vspace{0.6cm}

\noindent\large{\textbf{Mun-Hyok Ri\textit{$^{a}$}, Yong-Man Jang\textit{$^{b}$}, Chol-Jun Yu$^{\ast}$\textit{$^{a}$}, and Song-Un Kim\textit{$^{b}$}}}\vspace{0.5cm}

\noindent\textit{\small{\textbf{Received Xth XXXXXXXXXX 20XX, Accepted Xth XXXXXXXXX 20XX\newline
First published on the web Xth XXXXXXXXXX 200X}}}

\noindent \textbf{\small{DOI: 10.1039/b000000x}}
\vspace{0.6cm}

\noindent \normalsize{The structures of some bisphosphonates (clodronate, etidronate, pamidronate, alendronate, risedronate, zoledronate, minodronate) were obtained and analyzed, and their adsorption energies onto hydroxyapatite (001) surface were compared to find out ranking order of binding affinity, which shows that the adsorption energy is the largest for pamidronate, followed by alendronate, zoledronate, clodronate, ibandronate, the lowest for minodronate and etidronate.}
\vspace{0.5cm}
 \end{@twocolumnfalse}
  ]

\section{Introduction\label{sec:1}}

\footnotetext{\textit{$^{a}$~Department of Computational Materials Design (CMD), Faculty of Materials Science, Kim Il Sung University, Ryongnam-Dong, Taesong-District, Pyongyang, Democratic People's Republic of Korea. }}
\footnotetext{\textit{$^{b}$~Department of Organic Chemistry, Faculty of Chemistry, Kim Il Sung University, Ryongnam-Dong, Taesong-District, Pyongyang, Democratic People's Republic of Korea. }}

Bisphosphonates (BPs) are now widely used for the treatment of metabolic bone diseases including osteoporosis, Paget's disease and bone metastases\cite{Bartl2007,Ebetino2011,Coxon2006,Russell2008,Russell2011}.
In addition, BPs are also useful as novel ligands for well-defined radioactive metal complexes that can be used both for bone scanning imaging and for therapeutic applications\cite{Volkert1999}.
BPs have high affinity for calcium ions and therefore bind strongly to the principal bone mineral, hydroxyapatite (HAP), where they are internalized by bone-resorbing osteoclasts and inhibit their function\cite{Kontecka2002,Fleich1998,Rogers2000,Roelofs2006}.

\begin{figure}[!ht]
\centering
\includegraphics[width=0.3\textwidth]{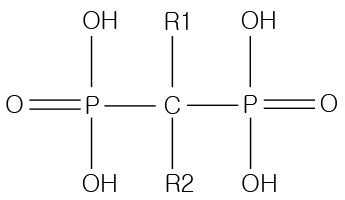}
\caption{General Structure of BPs}
\label{fig:BPstruct}
\end{figure}

\begin{figure*}[!ht]
\centering
\includegraphics[width=0.9\textwidth]{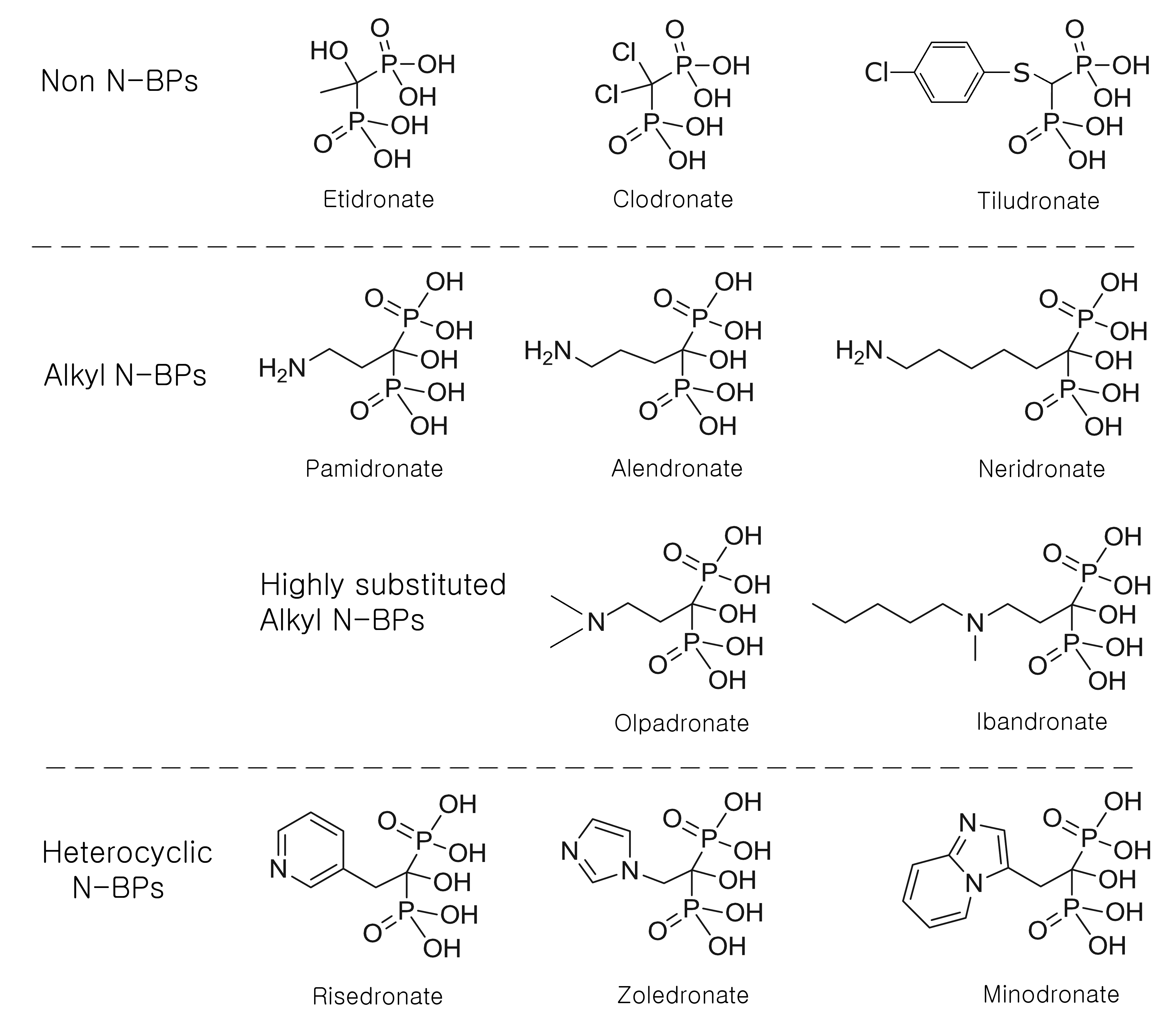}
\caption{Classification of BPs}
\label{fig:BPclass}
\end{figure*}

BPs are metabolically stable analogues of inorganic pyrophosphate (PPi), a naturally occuring modulator of calcification. Stability is confered by a carbon atom replacing the oxygen atom which connects two phosphonates.
BPs of medical interest all have two phosphonate groups sharing a common carbon atom between them(Figure~\ref{fig:BPstruct}).
BPs have chemical and physical properties similar to pyrophosphate\cite{Russell1970,Fleisch1970} but, due to their P-C-P backbone, they are considerably more resistant to heat and enzymatic hydrolysis.
Both phosphonate groups are required, as modifications to one or both reduce the affinity\cite{Ebetino1998}, as well as reduce biochemical potency\cite{Ebetino1995,Luckman1998}.
R1 substituents such as hydroxyl (OH) and amino (NH$_2$) can enhance chemical adsorption with their additional abilities to co-ordinate to calcium\cite{Benedict1982,Van1996}.
Varying R2 substituent(Figure~\ref{fig:BPclass}), the antiresorptive potency changes in several orders of magnitude\cite{Ebetino1998}.
Bisphosphonates with R2 chain containing a basic primary nitrogen atom in an alkyl chain (e.g., pamidronate, alendronate, and neridronate) are more potent antiresorptive than not containing nitrogen atom (e.g., clodronate and etidronate).
More highly substitution of nitrogen atom in an alkyl chain (e.g., olpadronate and ibandronate) can display further increase in an antiresorptive potency\cite{Papapoulos1988}.
The most potent antiresorptive bisphosphonates include those containing nitrogen atom within an heterocyclic ring (e.g., risedronate, zoledronate, and minodronate)\cite{Goa1998,Green1994}.
The overall pharmachological effects of BPs are mainly related to binding affinity for bone mineral (HAP) and inhibitory effects on osteoclasts\cite{Ebetino2011}.

Strong binding affinity of BPs on bone mineral can influence on some important biological properties of these drugs, including uptake and retention on bone, diffusion within bone, release from bone, and recycling back onto bone surface\cite{Oldfield2010,Russell2007}.
The mineral binding affinity may also affect the appropriate dosing interval and the persistence of effect after discontinuation of medication\cite{Black2007,Owens2007}.
Therefore, binding affinity of BPs on bone mineral may play an important role on the pharmacological effects of these drugs.

There are some reports on experiments of measuring and comparing binding affinities of BPs on bone.
By using a constant composition potentiostatic method, Nancollas et al studied the kinetic mineral binding affinities of bisphosphonates to identify zoledronate and alendronate as the higher affinity agents followed by ibandronate and risedronate\cite{Nancollas2006}.
Lawson and co-workers have developed FPLC (fast performance liquid chromatography) utilizing crystaline HAP, which showed that zoledronate had a longer retention time compared to risedronate, indicating a stronger binding affinity for HAP\cite{Lawson2009}.
Leu and co-workers studied bisphosphonate adsorption on human bone by competitive binding assays of radio-labeled bisphosphonates, which showed that most tested BPs, including etidronate, ibandronate, pamidronate, alendronate, risedronate, and zoledronate had comparable affinities, but tiludronate and especially clodronate displayed significantly weaker affinity for bone\cite{Leu2006}.
Jahnke et al, using NMR techniques, provided even more accurate comparison of thermodynamic binding affinities\cite{Jahnke2010} that parallels the Leu rankings\cite{Leu2006}. 
Mukherjee et al studied the thermodynamic properties of bisphosphonates binding to human bone in more detail using $^{31}$P NMR to obtain another set of affinity rankings\cite{Mukherjee2008-1,Mukherjee2008-2}, which showed that similar to the results of Nancollas et al\cite{Nancollas2006} risedronate has lower affinity, while zoledronate and alendronate have the highest affinity.
In recent studies, mass spectrometry (MS) has been applied for specific identification and quantification of BPs, which has enabled the efficient separation and quantification of BPs with their binding affinities\cite{Duan2010}.
The combination of FPLC with MS provides an accurate, precise, and robust method for quantitative analysis\cite{Duan2010}.
Although there are some differences in ranking orders of BPs binding affinity among various adsorption assays, in general, amino-alkyl BPs including pamidronate, alendronate, and neridronate have the highest affinities, while clodronate has the lowest affinity, and risedronate displays intermediate affinity (Table~\ref{tab:BPexp}).
\begin{table}[!ht]
\small
\caption{Comparison of rank orders of HAP binding affinities of BPs among different adsorption assays}
\label{tab:BPexp}
\begin{tabular*}{0.45\textwidth}{@{\extracolsep{\fill}}llllll}\hline
Ranking order & Exp$^a$ & Exp$^b$ & Exp$^c$ & Exp$^d$ & Exp$^e$ \\
\hline
1 & ZOL & ALN & PAM & PAM & PAM \\
2 & PAM & ZOL & ALN & ALN & ALN \\
3 & ALN & PAM & ZOL & ETD & NER \\
4 & IBN & RIS & RIS & NER & ETD \\
5 & RIS & ETD & IBN & ZOL & ZOL \\
6 & ETD & IBN & & RIS & IBN \\
7 & CLO & TIL & & IBN & MIN \\
8 & & CLO & & MIN & RIS \\
9 & & & & CLO & CLO \\
\hline
\end{tabular*}
$^a$ Constant composition kinetic studies of HAP crystal growth (Nancollas\cite{Nancollas2006}) \\
$^b$ Competitive binding assay of radio-labeled BPs (Leu\cite{Leu2006}) \\
$^c$ NMR-based competitive binding assay (Jahnke\cite{Jahnke2010}, Mukherjee\cite{Mukherjee2008-1,Mukherjee2008-2}) \\
$^d$ Fluorescence competitive binding assay (Duan\cite{Duan2010}) \\
$^e$ HAP FPLC (Duan\cite{Duan2010})
\end{table}

Structure-activity data in these studies show that small differences in BP structure lead to substantial changes in the three dimensional (3-D) shape and atomic orientation, resulting in significant changes in bone affinity.

Therefore, computational modeling and simulations are widely used to establish accurate and precise relationship between structure and bone affinity of BPs based on molecular and atomic level for development of  further promising BPs\cite{Lawson2005,Ebetino2005,Duarte2010}.
Modeling studies by Lawson et al\cite{Lawson2005} indicated that N atoms in the BP side chains can coordinate with OH groups in HAP with bonding efficiencies related to their overall binding affinity.
Comparative modeling of BPs by Ebetino et al\cite{Ebetino2005} demonstrated that N atoms of BPs can form a N-H-O hydrogen bond to the labile OH and a bifurcated interaction at the P-O oxygen atoms of HAP (001) surface.
Duarte et al\cite{Duarte2010} performed molecular mechanics simulations for molecular structures of 18 novel hydroxyl- and amino-BPs to examine the interaction between BPs and HAP and to extract relating structural characteristics of BPs and their affinities for the mineral, which are in agreement with in vitro and in vivo studies for some of the studied BPs.
These modeling and simulations are based on molecular mechanics with well-defined forcefields, that is not good for physico-chemical process such as adsorption of BPs on HAP surface accompanied with charge movement, in which density functional theory (DFT) provides reasonable results.

In recent years, DFT method are widely used for structural characteristics of BPs and HAP and interaction between them\cite{Zhu2004,Barrios2010,Ri2016}.
The surface energetics of HAP crystalline surfaces using ab initio density functional theory (DFT) calculations within the generalized gradient approximation (GGA) for the exchange-correlation functional has been studied by Zhu and Wu\cite{Zhu2004}, testing the effects of slab thickness, vacuum width between slabs, and surface relaxation on surface energy. Barrios\cite{Barrios2010} has investigated the interaction between collagen protein and HAP surface by using a combination of computational techniques, DFT and classical MD methods. 
More recently, the adsorption process of zoledronate on HAP (001) surface were examined in detail, which showed that the significant charge movement between BPs and HAP (001) surface occurs and the strong binding affinity of zoledronate with HAP is due to structural similarity\cite{Ri2016}.

In this paper, using ab initio DFT method, the molecular structures of 8 BPs (clodronate, etidronate, pamidronate, alendronate, ibandronate, risedronate, zoledronate, and minodronate) are obtained, and the binding affinity is evaluated via their adsorption energies onto HAP (001) surface.

\section{Computational Method\label{sec:2}}

DFT calculations are performed together with molecular mechanics (MM) calculations.
MM is less expensive computationally, but not more accurate and precise than DFT.
Therefore, MM is needed to find rough and globally-optimised structures using simulated annealing\cite{Kirkpatrick1983}, in which the temperature is increased from 300K to 5000K, and decreased back to 300K in 100,000 steps.
For MM calculations, we have used GULP (General Utility Lattice Program) code\cite{Gale2003} with Dreiding forcefield\cite{Mayo1990} and QEq atomic charges\cite{Rappe1991}.
Globally-optimised rough structure obtained in MM calculation goes for more accurate and precise structural relaxation in DFT calculation.

For DFT calculations, we have employed SIESTA code\cite{Soler2002} which solves numerically Kohn-Sham equation within DFT\cite{Hohenberg1964,Kohn1965} using a localized numerical basis set, namely pseudo atomic orbitals, and pseudopotentials for describing the interaction between ionic core (nucleus plus core electrons) and valence electrons. The BLYP GGA functional (the Becke exchange functional\cite{Becke1988} in conjunction with the Lee-Yang-Parr correlation functional\cite{Lee1988}) is used for exchange-correlation interaction between electrons. For all the atoms, Troullier-Martins\cite{Troullier1991} type norm-conserving pseudopotentials are generated within local density approximation (LDA)\cite{Perdew1981}, and the DZP type (double f plus polarization) basis sets are used. The mesh size of grid, which is controlled by energy cutoff to set the wavelength of the shortest plane wave represented on the grid, has taken a value of 200 Ry. Non-fixed atoms are allowed to relax until the forces converge less than 0.02 eV/$\text{A}^2$.

To simulate adsorption of BPs onto HAP surface, initial configuration of BPs adsorbed onto HAP surface must be given, from which globally-optimized configuration is obtained using simulated annealing in MM and further relaxation in DFT.

The adsorption energy of BPs onto HAP surface is given by
\begin{equation}
E_\text{ads}=E_\text{system}-(E_\text{HAP}+E_\text{BP}),
\label{eq:1}
\end{equation}
where $E_\text{HAP}$ is the energy of surface slab, $E_\text{BP}$, the energy of isolated BP molecule, and $E_\text{system}$, the total energy of system consisted of HAP surface slab and BP molecule adsorbed on it.

\section{Results and Discussion\label{sec:3}}
Table~\ref{tab:BPstruct} provides structural parameters of globally-optimized BPs structures, in which O(H) means oxygen atom linked to hydrogen atom, and P=O means double bond between phosphorous atom and oxygen atom. Bond length of C--R2 depends on R1 largely and bond length of P--C and bond angle of P--C--P vary significantly with R2, while other parameters remain almost unchanged, from which P--C--P structure is shown to be mainly related with binding affinity.
Also, $\pi$--bond of P=O reduces bond length compared to P--O(H).
\begin{table*}[!ht]
\small
\caption{Structural Parameters of BPs. P--O(H) means single bond between phorous atom and oxygen atom linked to hydrogen atom, and P=O means double bond between phorous atom and oxygen atom.}
\label{tab:BPstruct}
\begin{tabular*}{0.85\textwidth}{@{\extracolsep{\fill}}lrrrrrrrrr}
\hline
& CLO & ETD & PAM & ALN & IBN & RIS & ZOL & MIN \\
\hline
Bond Length & \multicolumn{8}{c}{} \\
\cline{1-1}
P--O(H) & 1.62 & 1.64 & 1.64 & 1.64 & 1.63 & 1.64 & 1.63 & 1.63 \\
P=O & 1.49 & 1.49 & 1.49 & 1.48 & 1.49 & 1.48 & 1.49 & 1.49 \\
P--C & 1.88 & 1.99 & 1.94 & 1.93 & 1.90 & 1.95 & 1.97 & 1.90 \\
C--R1 & 1.78 & 1.40 & 1.40 & 1.41 & 1.41 & 1.39 & 1.40 & 1.41 \\
C--R2 & 1.78 & 1.55 & 1.56 & 1.56 & 1.54 & 1.58 & 1.57 & 1.58 \\
\hline
Bond Angle & \multicolumn{8}{c}{} \\
\cline{1-1}
(H)O--P--O(H) & 97.0 & 96.9 & 96.4 & 95.4 & 99.2 & 95.1 & 97.3 & 102.1 \\
O=P--O(H) & 119.5 & 116.8 & 116.9 & 118.2 & 116.5 & 118.8 & 116.5 & 116.1 \\
C--P--O(H) & 105.1 & 110.4 & 109.6 & 106.9 & 106.1 & 107.0 & 110.3 & 105.2 \\
C--P=O & 108.9 & 105.8 & 107.0 & 109.9 & 111.3 & 108.5 & 105.7 & 111.0 \\
P--C--P & 102.4 & 93.7 & 94.3 & 98.6 & 96.9 & 101.1 & 90.9 & 94.6 \\
P--C--R1 & 110.2 & 110.9 & 113.2 & 107.7 & 109.0 & 113.2 & 112.9 & 109.1 \\
P--C--R2 & 110.1 & 113.8 & 112.8 & 115.7 & 113.0 & 108.6 & 114.9 & 114.7 \\
R1--C--R2 & 113.4 & 112.8 & 110.2 & 110.6 & 114.6 & 111.4 & 109.9 & 112.8 \\
\hline
\end{tabular*}
\end{table*}

\begin{figure}[!ht]
\centering
\includegraphics[width=0.45\textwidth]{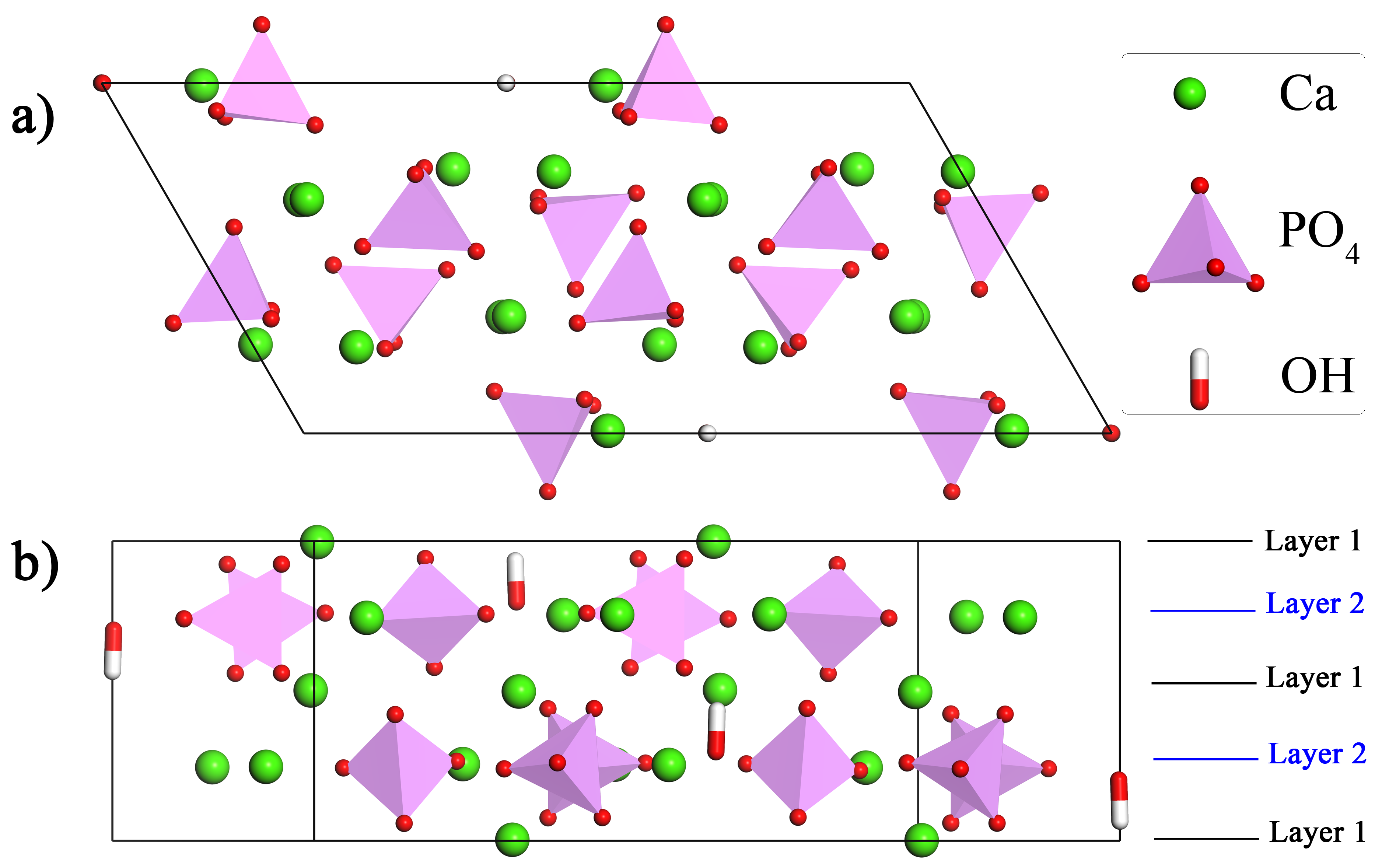}
\caption{Crystal Structure of Hydroxyapatite, a) Top View, b) Side View.}
\label{fig:hap}
\end{figure}

Structural and lattice parameters of HAP crystal and surface are provided in Table~\ref{tab:haplatt}, where HAP crystal is doubled along one axis normal to the orientation of hydroxyl group to keep its practical isotropy (See Figure~\ref{fig:hap}). For convenience, Ca atoms on Layer 1 are presented by Ca1, and those on Layer 2 by Ca2.
Comparison of structural paramters of BPs with those of HAP (Table~\ref{tab:haplatt}) shows that P--O(H) bond lengths of BPs  (1.62$\sim$1.64) are similar to those of HAP (1.578 for crystal, 1.583 for surface).
\begin{table}[!ht]
\small
\caption{Lattice Parameters of HAP crystal and surface}
\label{tab:haplatt}
\begin{tabular*}{0.45\textwidth}{@{\extracolsep{\fill}}lrr}
\hline
\multicolumn{3}{c}{Lattice Parameters} \\
\hline
a, b, c & \multicolumn{2}{c}{18.695, 9.354, 6.956} \\
$\alpha$, $\beta$, $\gamma$ & \multicolumn{2}{c}{90, 90, 119.86} \\
\hline
\multicolumn{3}{c}{Structural Parameters} \\
\hline
& Crystal & Surface \\
\hline
Average of bond lengths & & \\
\cline{1-1}
P--O & 1.578 & 1.583 \\
Ca1--O & 2.423 & 2.352 \\
Ca2--O & 2.553 & 2.434 \\
\hline
Average of bond angles & & \\
\cline{1-1}
O--P--O & 109.44 & 109.39 \\
O--Ca1--O & 98.62 & 101.50 \\
O--Ca2--O & 99.40 & 94.01 \\
\hline
\end{tabular*}
\end{table}

\begin{figure*}[!ht]
\centering
\includegraphics[width=0.95\textwidth]{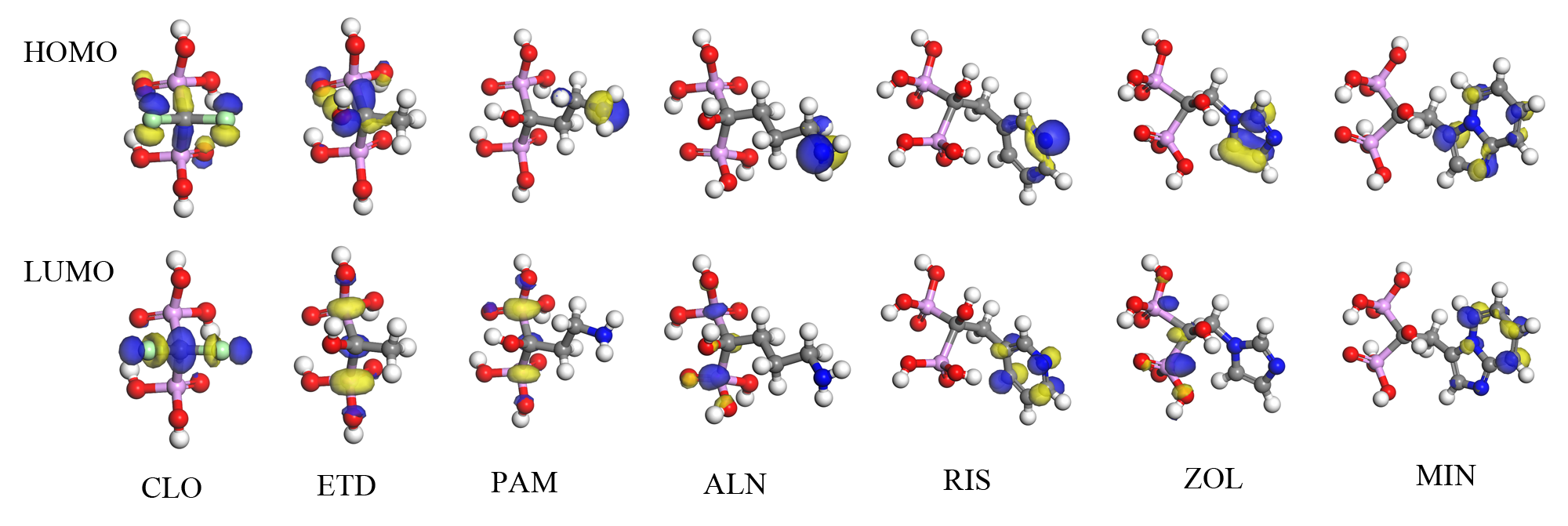}
\caption{HOMO and LUMO of BPs}
\label{fig:BPorbital}
\end{figure*}

Figure~\ref{fig:BPorbital} presents HOMO and LUMO of BPs which shows that HOMO is distributed around nitrogen atom for N-BPs and nitrogen atom behaves as electron donor in adsorption process.

Table~\ref{tab:BPhirshfeld} provides Hirshfeld charges~\cite{Hirshfeld1977} of BPs, in which O(=P) means oxygen atom double-bonding to phorous atom, and O(H) means oxygen atom linked to hydrogen atom. As shown in Table~\ref{tab:BPhirshfeld}, Hirshfeld charges of phosphorous atom and oxygen atoms double-bonding to former remain almost unchaged, and those of R1, R2, and carbon atom between two phosphorous atoms change significantly.
\begin{table*}[!ht]
\small
\caption{Hirshfeld Charges of BPs. O(=P) means oxygen atom double-bonding to phorous atom, and O(H) means oxygen atom linked to hydrogen atom.}
\label{tab:BPhirshfeld}
\begin{tabular*}{0.85\textwidth}{@{\extracolsep{\fill}}lrrrrrrrrr}
\hline
& CLO & ETD & PAM & ALN & IBN & RIS & ZOL & MIN \\
\hline
C & -0.025 & 0.045 & 0.038 & 0.025 & 0.014 & 0.034 & 0.032 & 0.007 \\
P & 0.490 & 0.498 & 0.494 & 0.494 & 0.500 & 0.496 & 0.500 & 0.501 \\
O(=P) & -0.336 & -0.353 & -0.355 & -0.358 & -0.334 & -0.343 & -0.342 & -0.335 \\
H & 0.152 & 0.157 & 0.154 & 0.162 & 0.142 & 0.160 & 0.153 & 0.147 \\
O(H) & -0.221 & -0.234 & -0.227 & -0.229 & -0.222 & -0.234 & -0.224 & -0.221 \\
R1 & 0.002 & -0.055 & -0.041 & -0.056 & -0.043 & -0.043 & -0.053 & -0.068 \\
R2 & 0.003 & 0.027 & 0.021 & 0.026 & 0.022 & -0.003 & -0.013 & 0.029 \\
\hline
\end{tabular*}
\end{table*}

\begin{figure*}[!ht]
\centering
\includegraphics[width=0.95\textwidth]{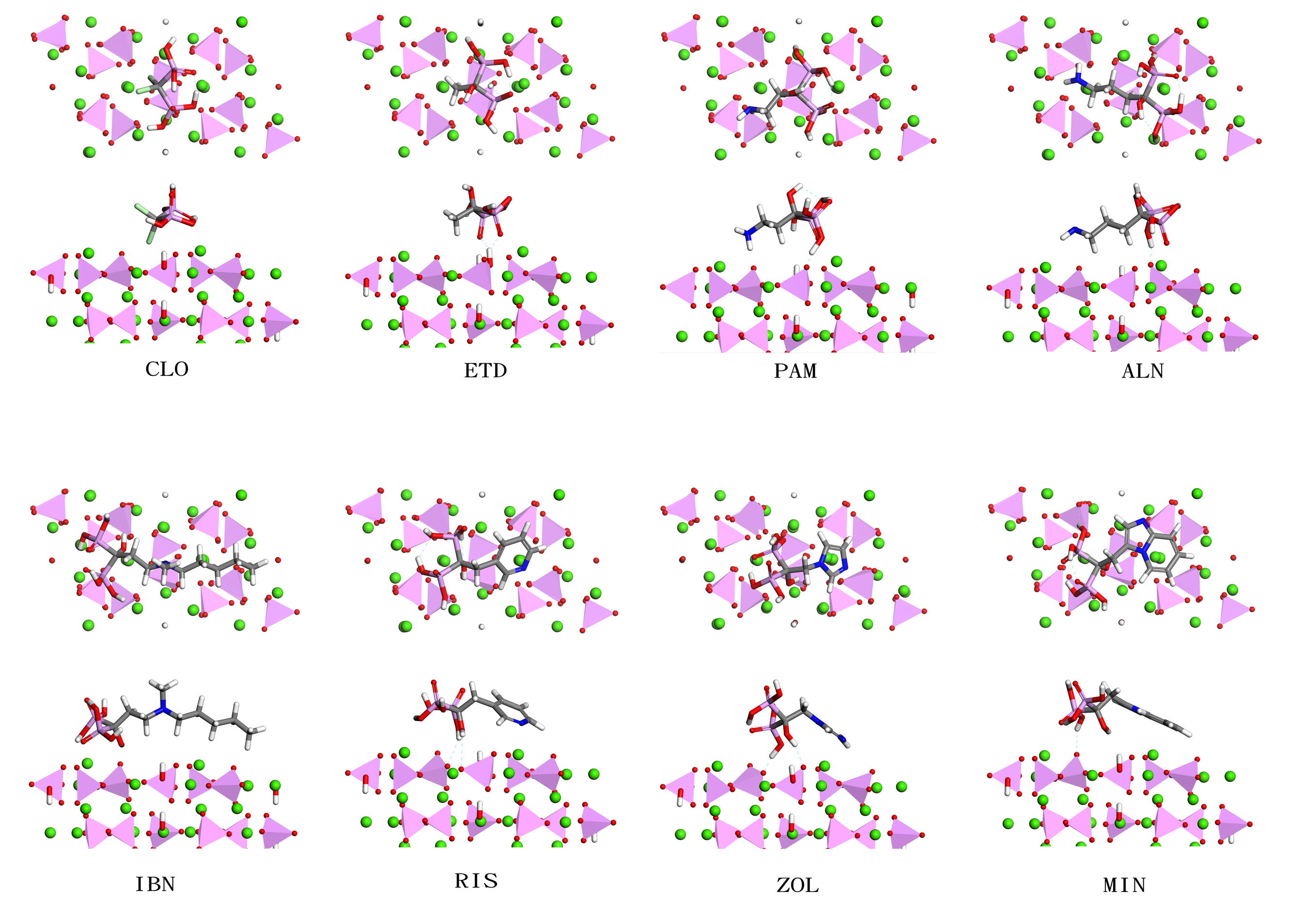}
\caption{Adsorption Geometries of BPs onto HAP (001) surface.}
\label{fig:ads}
\end{figure*}

Figure~\ref{fig:ads} shows the adsorption geometries of BPs onto HAP (001) surface, which indicates that hydrogen bond between OH of BPs and O of HAP plays an important role in strong binding of BPs onto hydroxyapatite.

Table~\ref{tab:haphirshfeld} shows Hirshfeld charges of HAP crystal and surface (before and after BPs adsorption). As shown in Table~\ref{tab:haphirshfeld}, atomic charges of surface differ remarkably from those of crystal before adsorption, while the adsorption of BPs relaxes atomic charges of HAP surface towards  those of crystal, so that the adsorption of BPs stabilizes HAP surface.
\begin{table*}[!ht]
\small
\caption{Hirshfeld Charges of HAP crystal and surface. Ca1 means calcium atom on Layer 1, Ca2 means calcium atom on Layer 2, O1, O2 and O3 mean oxygen atoms of phosphate tetrahedron, in which O1 places on Layer 1, and O2 and O3 on Layer 2.}
\label{tab:haphirshfeld}
\begin{tabular*}{0.95\textwidth}{@{\extracolsep{\fill}}lrrrrrrrrrr}
\hline
& & & \multicolumn{8}{c}{HAP Surface combined with BPs} \\
\cline{4-11}
Element & Crystal & Surface & CLO & ETD & PAM & ALN & IBN & RIS & ZOL & MIN \\
\hline
Ca1 & 0.429 & 0.783 & 0.555 & 0.616 & 0.592 & 0.553 & 0.550 & 0.516 & 0.532 & 0.524 \\
Ca2 & 0.415 & 0.508 & 0.460 & 0.478 & 0.479 & 0.471 & 0.471 & 0.468 & 0.477 & 0.478 \\
O1 & -0.290 & -0.377 & -0.358 & -0.347 & -0.335 & -0.340 & -0.348 & -0.349 & -0.341 & -0.351 \\
O2 & -0.283 & -0.318 & -0.314 & -0.308 & -0.310 & -0.310 & -0.310 & -0.313 & -0.304 & -0.312 \\
O3 & -0.296 & -0.312 & -0.309 & -0.290 & -0.307 & -0.308 & -0.307 & -0.307 & -0.307 & -0.309 \\
P & 0.533 & 0.481 & 0.494 & 0.501 & 0.502 & 0.502 & 0.498 & 0.499 & 0.503 & 0.497 \\
H & 0.130 & 0.128 & 0.120 & 0.130 & 0.130 & 0.130 & 0.130 & 0.131 & 0.136 & 0.135 \\
O(H) & -0.355 & -0.356 & -0.359 & -0.366 & -0.362 & -0.359 & -0.356 & -0.358 & -0.362 & -0.358\\
\hline
\end{tabular*}
\end{table*}

Table~\ref{tab:adsE} shows the adsorption energies of BPs onto HAP(001) surface, which indicates that PAM has the greatest value, followed by ALN, ZOL, CLO, IBN, RIS, MIN, and ETD, which  parallels the ranking orders of Table~\ref{tab:BPexp}.
\begin{table}[!ht]
\small
\caption{Adsorption Energies of BPs onto HAP(001) surface.}
\label{tab:adsE}
\begin{tabular*}{0.48\textwidth}{@{\extracolsep{\fill}}lrrrr}
\hline
& E$_\text{BP}$ (eV) & E$_\text{HAP}$ (eV) & E$_\text{BP-HAP}$ (eV) & E$_\text{Ads}$ (eV) \\
\hline
CLO & -3997.45 & -50669.94 & -54669.74 & -2.35 \\
ETD & -3840.24 & -50669.94 & -54512.35 & -2.17 \\
PAM & -4313.33 & -50669.94 & -54986.60 & -3.33 \\
ALN & -4500.17 & -50669.94 & -55173.33 & -3.21 \\
IBN & -5436.13 & -50669.94 & -56108.38 & -2.31 \\
RIS & -4931.92 & -50669.94 & -55604.11 & -2.25 \\
ZOL & -4876.41 & -50669.94 & -55549.28 & -2.92 \\
MIN & -5528.59 & -50669.94 & -56200.71 & -2.17 \\
\hline
\end{tabular*}
\end{table}
%

%
%

\section{Conclusions}
In this paper, the geometries of BPs were studied to find out that P--C--P structure is sensitive to R2 group and mainly correlated to binding affinity onto hydroxyapatite crystal.
For N-BPs, nitrogen atom is shown to behaves as donor during the adsorption through HOMO and LUMO analysis.

Hirshfeld charge analysis shows that during the adsorption process, charges of BPs and HAP change reamarkably, and BPs adsorption realaxes charges of HAP surface towards those of HAP crystal, so that it can stabilize HAP surface.

The adsorption geometries of BPs onto hydroxyapatite (001) surface shows that hydrogen bond between OH of BPs and O of HAP plays an important role in strong binding of BPs onto hdyroxyapatite.

Comparing the adsorption energies of BPs onto hdyroxyapatite (001) surface, obtained by first-principles calculation, pamidronate has the greatest value, and etidronate has the lowest value, which parellels the experimental ranking order, i.e. adsorption energy is the main factor determining the binding affinity of BPs onto hydroxyapatite.

\footnotesize{
\bibliography{reference} 
\bibliographystyle{plain} 
}

\end{document}